# Tuning the structural instability of SrTiO$_3$ by Eu doping: the phase diagram of Sr$_{1-x}$Eu$_x$TiO$_3$

Zurab Guguchia[1], Alexander Shengelaya[2], Hugo Keller[1], Jürgen Köhler[3], Annette Bussmann-Holder[3]

[1]Physik Institut der Universität Zürich, Winterthurerstr. 190, CH-8057 Zürich, Switzerland

[2]Department of Physics, Tbilisi State University, Chavchavadze 3, GE-0128 Tbilisi, Georgia

[3]Max-Planck-Institut für Festkörperforschung, Heisenbergstr. 1, D-70569 Stuttgart, Germany

The phase diagram of Sr$_{1-x}$Eu$_x$TiO$_3$ is determined experimentally by EPR and resistivity measurements and analyzed theoretically within the self-consistent phonon approximation as a function of x (0.03 ≤ x ≤ 1.0). The transition temperature of the structural instability of the system increases nonlinearly to higher temperatures with increasing x. This is interpreted theoretically by a substantial alteration in the dynamics caused by a change in the double-well potential from broad and shallow to narrow and deep.

SrTiO$_3$ and EuTiO$_3$ have a variety of aspects in common which might enlarge their field of applications as mixed crystals or layered materials enormously. The atomic radii of Sr and Eu, their lattice constants, and their valencies are identical in the perovskite ABO$_3$ structure. Both compounds show a strong tendency towards a ferroelectric instability signaled by a transverse optic long wave length mode softening, which is, however, suppressed by quantum fluctuations [1, 2, 3]. The extrapolated values of the transition temperatures T$_F$ are 37K for SrTiO$_3$ (STO) [4 – 6] and <-150K for EuTiO$_3$ (ETO) [2, 3]. In addition, very recent experimental and theoretical studies [7, 8] have demonstrated another commonality between these compounds, namely a structural phase transition at elevated temperatures which in STO has been demonstrated to be caused by the oxygen octahedral tilting instability whereas its precise nature is unknown in ETO. While in STO the transition is observed at $T_S$=105K, the one in ETO sets in at $T_S$=282K. Even though it remains speculative to associate this structural phase transition with the same octahedral tilting instability as in STO, the theoretical analysis of it supports this assumption. The large difference between both structural transition temperatures has motivated us to explore the phase diagram of Sr$_{1-x}$Eu$_x$TiO$_3$ as a function of x. The x-dependence of the low temperature antiferromagnetic transition of ETO at $T_N$=5.5K [9, 10] is not studied, even though substantial changes are expected with varying x.

For the end members of the mixed crystals we have shown [7, 8] that their dynamical behavior can be understood within the framework of the polarizability model [11 – 13]. Specifically we have demonstrated that the *same* set of parameters applies to both systems. The only difference is caused by the mass of the *A*-sublattice (in ATiO$_3$) which is enhanced in EuTiO$_3$ as compared to SrTiO$_3$ For the compounds with x=0.03, 0.25, 0.5, 0.75 we use again the *same* parameters and change the *A* mass according to the substitution level x. The double-well defining parameters, which characterize the rotational instability, have been taken as x-dependent averages of those of the end members. It is important to note here, that STO and ETO largely differ with respect to their local double-well potentials since the one of STO is broad and shallow while the one of ETO is narrow and deep [7, 8]. Typically these characteristics provide evidence for displacive dynamics being realized in STO, while order/disorder aspects are realized in ETO. Similarly, the coupling between the BO$_3$ units is taken as x dependent averages of the pure compounds. In the following we present results for the phase diagram of Sr$_{1-x}$Eu$_x$TiO$_3$ as determined by EPR, resistivity measurements and the above lattice dynamical calculations. From the data as well as the theoretical analysis it is concluded that a structural instability is present in all samples (most likely related to the oxygen ion rotational instability) which appears as a distinct anomaly in the experiments.

Samples of $Sr_{1-x}Eu_xTiO_3$ have been prepared analogous to the pure compounds as described in Ref. 7. The values of x are x=0.03, 0.25, 0.5, 0.75, 1. The polycrystalline samples have been studied by means of the electron paramagnetic resonance (EPR) technique with the emphasis on investigating and characterizing the structural instability in detail. EPR experiments were performed with a Bruker EMX spectrometer at $X$-band frequencies ($\nu \approx 9.4$ GHz) equipped with a continuous He gas-flow cryostat in the temperature range $4.2 < T < 300$ K. Here, however, we restrict the discussion to temperatures $T > 50$ K. The EPR method [14] is useful in the detection of structural phase transitions in perovskite oxides as has been demonstrated for the oxygen octahedral instability in STO where a broadening of the EPR line width of a $Fe^{3+}$-$V_O$ pair center at $T_S$ has been observed [15]. This almost divergent line width at $T_S$ has been explained in terms of spin-soft-phonon-mode coupling [16] where the spins reflect the temperature dependence of the soft mode. In the present experiment also the change of the EPR line width with temperature was studied. Opposite to the EPR study on STO where single crystal data have been used, in the present study powder samples were studied which do not admit a similarly detailed analysis of the data as has been done for STO [17]. The present study has, however, the advantage that the magnetic ion $Eu^{2+}$ is intrinsic and severs as a perfect target for EPR.

In Fig. 1a the EPR spectra of $Sr_{1-x}Eu_xTiO_3$ with x=0.03, 0.25, 05, 0.75, and 1 are shown at T=300K. For all x a weakly asymmetric broad resonance line is observed which can be well described by a Dyson shape [18 – 20]:

$$P(H) \propto \frac{\Delta H + \alpha(H - H_{res})}{(H - H_{res})^2 + \Delta H^2}, \tag{1}$$

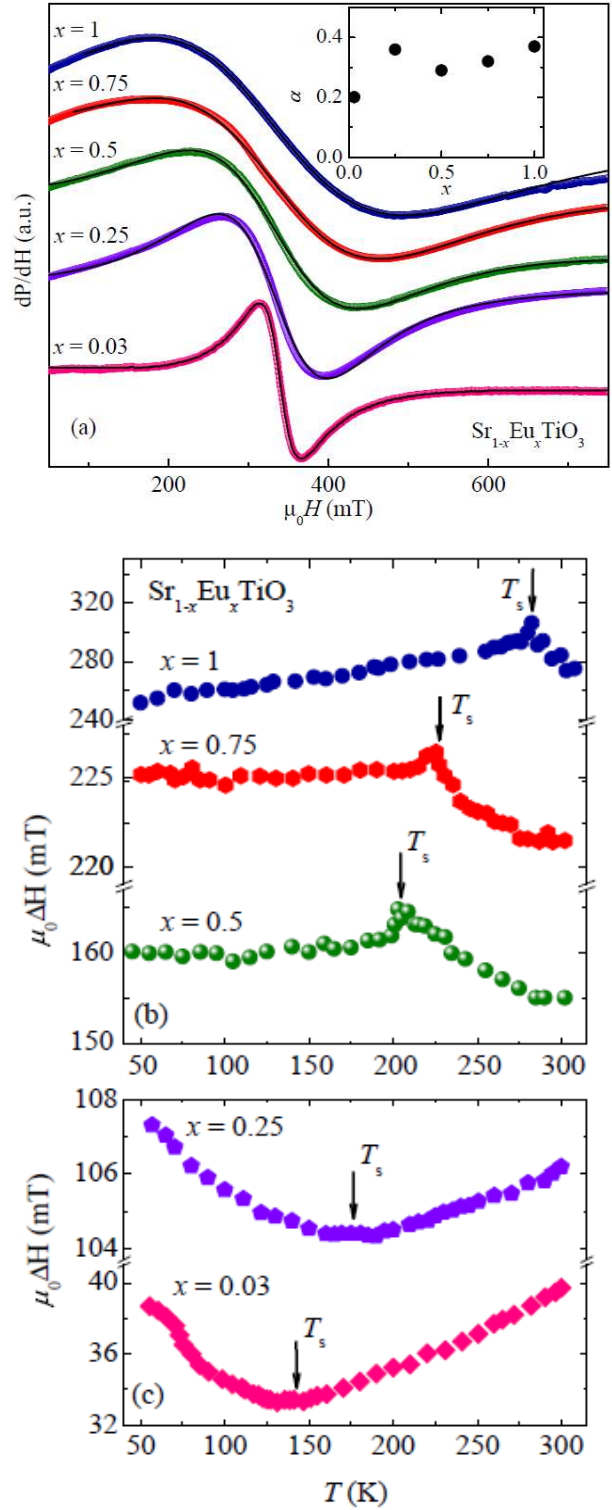

**Figure 1:** (a) (Color online) EPR spectra dP/dH of Eu$^{2+}$ in Sr$_{1-x}$Eu$_x$TiO$_3$ for 0.03≤x≤1 at T=300K. The experimental data are in color, the fit with Eq. (1) is given by the black lines. In the inset the x-dependence of α is shown. (b) (Color online) Temperature

dependence of the EPR line width $\mu_0\Delta H$ for $Sr_{1-x}Eu_xTiO_3$ samples with $x=0.5\leq x\leq 1$. The arrows indicate the structural phase transition temperature $T_S$. (c) (Color online) Temperature dependence of the EPR line width $\mu_0\Delta H$ for $Sr_{1-x}Eu_xTiO_3$ samples with $x=0.25, 0.03$. The arrows indicate the structural phase transition temperature $T_S$.

corresponding to a Lorentz line at a resonance field $H_{res}$ with half width at half maximum $\Delta H$ and a contribution $0 \leq \alpha \leq 1$ of dispersion to the absorption resulting in a characteristic asymmetry. The parameter $\alpha$ results from a mixture of absorptive and dispersive components of the susceptibility. This is caused by the non-uniform distribution of the microwave electromagnetic field. $\alpha$ depends on the sample size, geometry, and skin depth and its x-dependence is displayed in the inset to Fig. 1a. If the skin depth is small in comparison to the sample size, $\alpha$ approaches 1. However, here we have used samples in powder form in order to obtain a more intense signal, whereby the grain size is comparable to the skin depth. From Fig. 1a it is obvious that the EPR signal broadens with increasing Eu content, which signals the increasing magnetic dipolar interaction between the $Eu^{2+}$ ions with decreasing Eu – Eu distance. The Lorentzian line shape is a signature of the exchange narrowing process due to strong exchange coupling between the magnetic $Eu^{2+}$ ions. The fine and hyperfine structures, which are expected for the single $Eu^{2+}$ ions, are not observed in our samples. This implies that starting from the smallest Eu concentration $x = 0.03$ studied here, the exchange narrowing process is sufficient to merge the entire spectrum into a single EPR line. The line width caused by the dipole-dipole interaction for $EuTiO_3$ [21] is calculated following the theory of exchange narrowing of Anderson and Weiss [22] where the high-temperature limit of the line width $\Delta H_\infty$ can be estimated as

$$\Delta H_\infty = \frac{\hbar}{g\mu_B} \frac{\langle v_{DD}^2 \rangle}{v_{ex}} \tag{2}$$

where $v_{ex}$ is the exchange frequency between the Eu spins and $v_{DD}^2$ denotes the second moment of the resonance frequency distribution due to the dipolar interaction and reads

$$\langle v_{DD}^2 \rangle = g^4 \mu_B^4 \frac{3S(S+1)}{2h^2} \sum_{j \neq i} \frac{1+\cos^2\theta_{ij}}{r_{ij}^6}, \tag{3}$$

where $r_{ij}$ and $\Theta_{ij}$ denote the distance between spin $i$ and $j$ and the polar angle of the external magnetic field with respect to the direction of $r_{ij}$. The above relation is valid when the exchange coupling between the spins is larger than the Zeeman energy and has been derived , e.g., in Ref. 23. This condition is certainly fulfilled in the present case. The main contribution results from the four nearest Eu neighbors at $r_{ij} = a = 3.905$ Å. Since powder samples were measured, we assume an average value $\langle 1+\cos^2\theta \rangle = 4/3$. With g = 2 and S = 7/2 one obtains $\langle v_{DD}^2 \rangle = 96.2 GHz^2$.

The exchange coupling $J_{Eu}$ between the $Eu^{2+}$ ions is determined from the Curie-Weiss temperature $T_N = 5.5K$ using the Weiss molecular-field equation $3k_B T_N = J_{Eu} z S(S+1)$ with z = 4 exchange-coupled nearest neighbors as $J_{Eu}/k_B \approx 0.26K$. Then the exchange frequency can be approximately estimated by $(hv_{ex})^2 \approx zS(S+1)$ resulting in $v_{ex} \approx$ 41.25 GHz. Thus the line width due to dipolar broadening is estimated to be $\mu_0 \Delta H_\infty = 84 mT$. This value is considerably smaller than observed experimentally for ETO: $\mu_0 \Delta H \approx 270 mT$ at room temperature. However, in agreement with Ref. 21, the dependence of $\Delta H$ on x is linear. As will be shown below and already explained previously in the context of the line width increase in $CrBr_2$ [22], strong spin-lattice coupling is the origin of this anomaly. In Fig. 1b the temperature dependence of the line width $\Delta H$ is shown which changes qualitatively with increasing Eu content. While for x=0.03 and 0.25 the line width decreases with decreasing temperature to reach a minimum at $T_S$ and increase below this again, for the remaining samples (x=0.5, 0.75, 1), an increase in the line width is seen which reaches a maximum at the temperature $T_S$ followed by a smooth decrease. These distinctly different temperature dependencies suggest that a crossover from metallic (x≤0.25) to semiconducting (x≥0.5) behavior takes place between x=0.25 and x=0.5. The metallic properties of low doped STO might be astonishing since pure STO is a large gap insulator. However, it is well known that n-doped STO rapidly becomes a semiconductor and even superconducting [23 – 27], whereas Nb-doped STO is metallic and superconducting for very small Nb doping concentrations [28]. Since Eu easily changes its valency from 2+ to 3+, small amounts of $Eu^{3+}$ can give rise to the observed metallic properties of the samples.

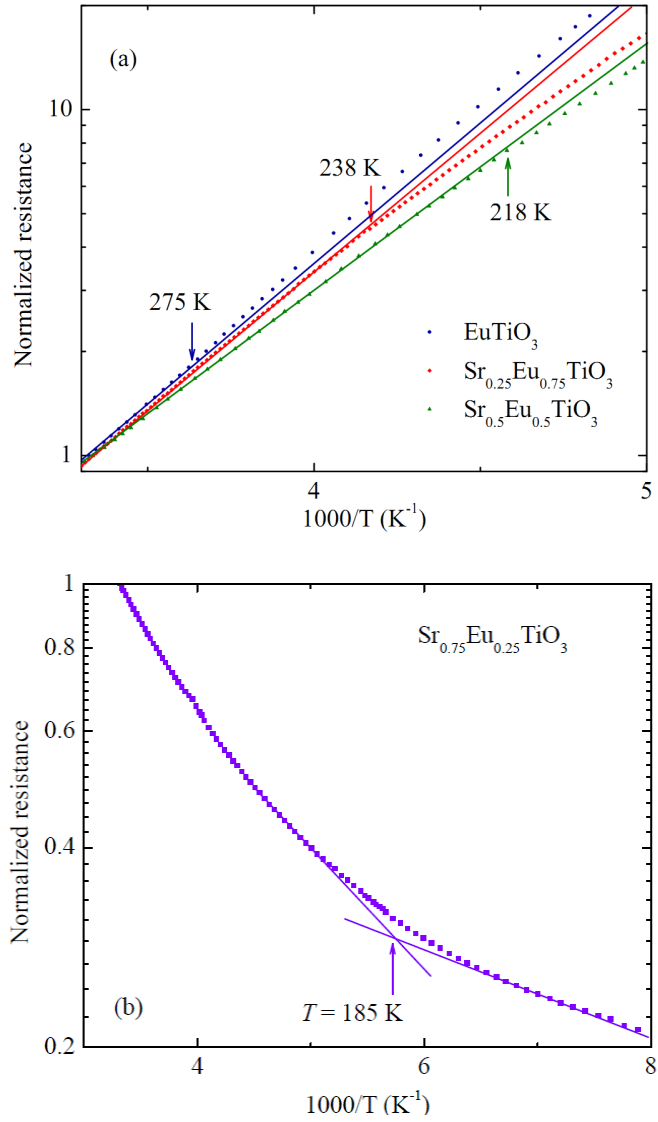

**Figure 2:** (Color online) a) Normalized resistance $\rho/\rho_0$ as a function of inverse temperature with $\rho_0 = \rho(T=300K)$ of $Sr_{1-x}Eu_xTiO_3$ for x ≥ 0.5. b) shows the same as a) for the sample with x=0.25.

In metallic samples the EPR line width follows an empirical relation $\Delta H = a + bT$ where both *a* and *b* are material dependent constants and generally positive. *b* is determined by thermal fluctuations of the exchange interaction of localized moments with the conduction electrons (Korringa relaxation) [29], whereas the residual line width *a* stems from spin-spin interactions of localized moments and lattice defects. A more microscopic expression of the empirical rule has been derived by Huber and Seehra [30] who determined the line width temperature dependence as:

$$\Delta H = \frac{\hbar(C + f(\varepsilon))}{g\mu_B T\chi} \tag{4}$$

where $g$ is the electronic $g$ factor, $\mu_B$ the Bohr magneton and $C$ and $f(\varepsilon)$ ($\varepsilon = (T - T_C)/T_C$) are the non-critical and critical contributions to $\Delta H$. While the former contribution leads to a T independent line width, the latter becomes important in the vicinity of the magnetic transition temperature $T_N$ only. The data presented in Fig. 1b clearly demonstrate that a temperature dependence of $\Delta H$ is present above the critical regime quite analogous to $CrBr_3$. For this system Huber and Seehra [30] have explained the additional temperature dependence in terms of spin-phonon coupling by extending the spin-spin Hamiltonian by the term:

$$H_{sp} = A_1\left[3S_z^2 - S(S+1)\right]Q_3 + \sqrt{3}A_2\left[S_x^2 - S_y^2\right]Q_2 \tag{5}$$

taking into account only coupling to $\Gamma_3$ vibrations. Here $S$ is the total spin of the ion under consideration with $S_i$ ($i = x, y, z$) being the Cartesian components of the spin. $A_i$ are coupling constants and $Q_i$ the phonon normal coordinates. While in cubic symmetry $A_1 = A_2$, in the tetragonal symmetry $A_1 \neq A_2$. This implies that a structural phase transition directly affects the EPR line width and induces pronounced changes in it upon symmetry lowering. In addition, the normal mode coordinates adopt a temperature dependence in the presence of a soft mode as anticipated for the oxygen octahedral rotational mode, which we assume to cause the structural anomaly [16]. It is important to emphasize that the expected soft mode remains temperature dependent not only for $T>T_S$ but also for $T<T_S$ where it hardens according to the Curie law. The temperature dependence of $\Delta H$ reflects this dependence extremely well for samples with x=0.03 and 0.25 where the line width follows approximately the Curie law. For samples with x≥0.5 the temperature dependence is reversed as compared to the two low doped samples, but still exhibiting an anomaly at $T_S$. This qualitative change is a consequence of the change in the conductivity (see below) which moves from conducting to semiconducting between x=0.25 and 0.5 (see below). In semiconducting or insulating samples the EPR line width is also dependent on the spin-phonon coupling through the local crystal field potential $D$ [16] which may change during a phase transition. As has been shown by Owen [16], the zero field splitting $D$ can be expressed in terms of the order parameter $\Phi$ related to the rotational instability like $D = C_2 \Phi^2$, with $C_2$ being the Landau coefficient in the expansion of the free energy in $\Phi$. The derivation of the zone boundary soft mode $\omega_{TA}(q = 2\pi/a)$ (see below) shows that this follows mean-field behavior, namely, $\omega_{TA}^2(q = 2\pi/a) \approx (T - T_S)$, $D \approx (T - T_S)$ as well. In this case $\Delta H$ diverges like $(T - T_S)^{-1/2}$ at $T_S$. The data presented in Fig. 1b for x ≥0.5 are in accord with such a coupling to the rotational order parameter, with the divergence being diminished by dilution and or impurities.

Even though the EPR data can not give direct evidence for the rotational instability of the oxygen ion octahedra analogous to STO, the coincidence of the maximum in the EPR line width in ETO with the specific heat anomaly, let us to conclude that the same structural phase transition occurs here and also in the doped samples.

The origin of the different behaviors of samples with x≤0.25 and x≥0.5 was further investigated by performing resistivity measurements. The data for samples with x≥0.25 are shown in Fig. 2a, and Fig. 2b corresponds to the data taken for x=0.25. All data have been normalized to their values at T=300K and are plotted logarithmically versus the inverse temperature in order to highlight their semiconducting properties. Obviously, a change from metallic to semiconducting behavior sets in for x≥0.5 where the resistivity $\rho$ follows a semiconducting behavior $\rho = \rho_0 \exp(-\Delta/kT)$ with the semiconducting gap $\Delta$ changing at $T_S$ as indicated by arrows in the figure. The semiconducting gaps above and below the transition temperature are given in Table I together with the resistivity values at T=300K and 120K.

**Table I:** The values of the semiconducting gaps $\Delta$ of $EuTiO_3$, $Sr_{0.25}Eu_{0.75}TiO_3$, and $Sr_{0.5}Eu_{0.5}TiO_3$ for temperatures $T>T_S$ and $T<T_S$ in eV and the values of the resistivity $\rho$ for $Sr_{1-x}Eu_xTiO_3$ samples with x=0.25, 0.5, 0.75, 1 at T=300K and T=120K.

|  | $EuTiO_3$ | $Sr_{0.25}Eu_{0.75}TiO_3$ | $Sr_{0.5}Eu_{0.5}TiO_3$ | $Sr_{0.75}Eu_{0.25}TiO_3$ |
|---|---|---|---|---|
| **$\Delta$ (eV) $T>T_S$** | 0.63 | 0.44 | 0.46 | |
| **$\Delta$ (eV) $T<T_S$** | 0.71 | 0.454 | 0.48 | |
| **$\rho(T = 300K)$ ($\Omega$)** | 586.59 | 55391.93 | 22313.67 | 0.01952 |
| **$\rho(T = 120K)$ ($\Omega$)** | $1.3 \times 10^6$ | $1.9 \times 10^7$ | $2 \times 10^6$ | $3.7 \times 10^{-3}$ |

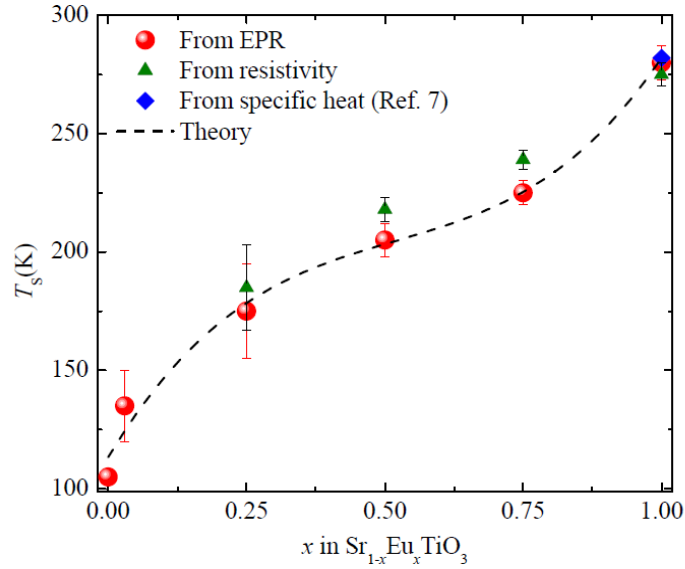

**Figure 3:** The phase diagram of $Sr_{1-x}Eu_xTiO_3$ as determined from the line width broadening (Fig. 1), the resistivity data (Fig. 2), specific heat [7], and calculated theoretically (dashed line).

The sample with x = 0.25 shows a metallic resistivity and exhibits a crossover point at $T_S$ where the low and high temperature linear dependencies intercept. Note, that the latter data have been plotted with the same convention as the former ones in order to compare the results directly. The temperatures $T_S$ at which line width anomalies are seen are identical to those temperatures where $\Delta$ changes, respectively the intercept in the resistivity appears. From both data, EPR and resistivity measurements, it is thus possible to construct a consistent phase diagram for $Sr_{1-x}Eu_xTiO_3$ which is shown in Fig. 3.

As is obvious from Fig. 3, the structural instability is nonlinearly dependent on the Eu content, rather unexpectedly when taking into account that the radii of Eu and Sr and the lattice constants of the end members are identical. The nonlinear behavior can thus not be a consequence of any lattice mismatch but must be inherent and either stemming from the mass difference of the two ions or be of lattice dynamical origin or a consequence of both together. From our previous analysis of the structural instability of the end member compounds [7, 8], we have observed that the self-consistently derived double-well potentials of them differ grossly, since the one of STO is shallow and broad (reminiscent of a purely displacive transition) whereas the one of ETO is deep and narrow, as is typically observed for order/disorder driven phase transitions in spite of the fact that all other model parameters remain the same. Since the soft mode dynamics of the mixed crystals are experimentally not known, we use for them the same parameters as before for the end members and employ x-dependent averages of the double-well defining parameters $g_2$ and $g_4$, where $g_2$ is the attractive electron-ion interaction parameter and $g_4$ the fourth order repulsive term. Similarly the *A* sublattice mass is determined as an x-dependent average of Sr and Eu. The resulting double-well potential barrier heights given in terms of $g_2/g_4$ are shown in Fig. 4a. Obviously, for x>0.25 the barrier height becomes x-independent, while for x≤0.25 a strong x-dependence is observed. This variation of the barrier height can clearly not be attributed to the simultaneous changes in the resistivity and the EPR line width but must be of dynamical origin where mass changes or crossover physics dominate. However, a definite conclusion can not be drawn here as long as the corresponding experiments have not been carried through. From the potential parameters the zone center soft mode has been calculated as a function of x which – in spite of obvious similarities between STO and ETO – is also distinctly different in both systems since it extrapolates to zero at finite temperature in the STO [4 – 6] while it is far in the negative temperature scale for ETO [2, 3]. The results for all x are shown in Fig 4b. Interestingly, the soft zone center mode shows an enormous dependence on x for x≤0.25 where it shifts considerably to higher values with increasing x. However, for x≥0.5 the x-dependence has vanished and all three curves fall on a single line with the same zero point extrapolated intercept.

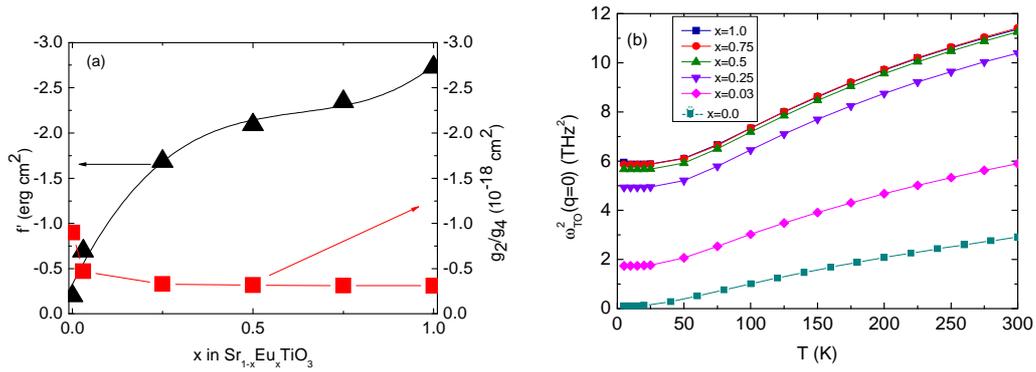

**Figure 4:** (Color online) (a) The x-dependence of the second nearest neighbor octahedral coupling $f'$ (triangles) and of the double-well potential height $g_2/g_4$ (squares) $Sr_{1-x}Eu_xTiO_3$. (b) Temperature and x-dependence of the squared soft optic mode frequency $\omega_{TO}^2$ for $Sr_{1-x}Eu_xTiO_3$

For SrTiO$_3$ it has been shown, that the zone center and the zone boundary instabilities are interrelated with each other through polarizability effects [31]. The same interrelation should naturally also be present in the mixed crystals. With the double-well defining parameter values and the temperature dependence of the long wave length optic mode frequencies it is possible to deduce the temperature dependence of the zone boundary related acoustic mode energy corresponding to the octahedral rotational instability. This alone is, however, not enough to reproduce the experimental data for $T_S$. It is necessary to also correct for the spin-phonon coupling which – as has been shown previously [7] – strongly suppresses the zone boundary acoustic mode frequency. We infer this correction indirectly by modifying the second nearest neighbor coupling $f'$ [7, 8] such as to yield the correct $T_S$. Interestingly, we find that this coupling follows the x dependence of $T_S$ (Fig. 4a) and increases with increasing x from almost zero to $|f'|= 2.9$. This strong increase in $|f'|$ has not only the consequence that the zone boundary instability at $T_S$ moves to higher temperatures but also that the acoustic mode for small momentum is lowered in energy whereby the optic-acoustic mode coupling is suppressed in this momentum range. This – in turn – stabilizes the elastic constants already for $x \geq 0.25$ while for x=0.03 and STO these are very soft.

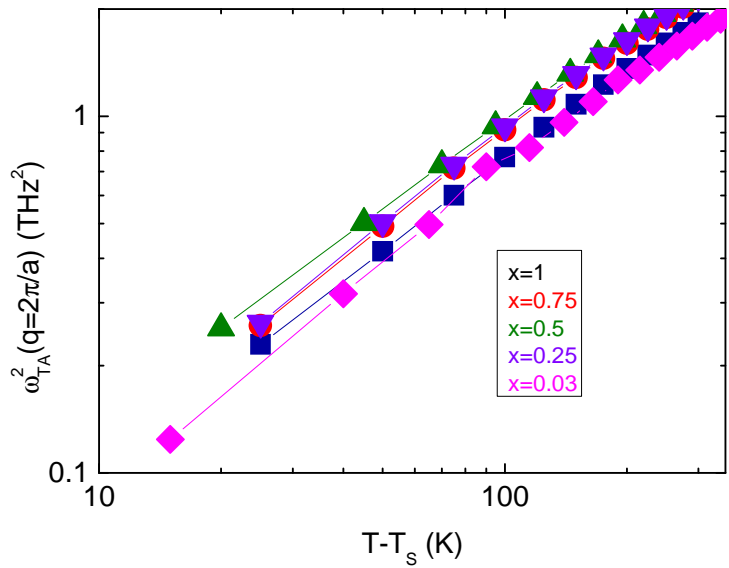

**Figure 5:** (Color online) Double logarithmic plot of the squared zone boundary soft mode of Sr$_{1-x}$Eu$_x$TiO$_3$ as a function of temperature.

The calculated zone boundary frequency responsible for the structural phase transition is shown in Fig. 5 as a function of $(T-T_S)$ for the x-values discussed in this paper. Interestingly, no clear distinction between the different x-values can be established. All curves follow almost the same temperature dependence in a mean-field manner ( $\omega_{TA}^2(q=2\pi/a) \approx (T-T_S)^\gamma$, $\gamma=1$ ) and no qualitative changes are seen. This observation clarifies, that the origin of the crossover physics appearing in the mixed crystals can not be attributed to the structural instability but is exclusively triggered by polarizability effects. On the other hand, the mean-field behavior observed for the soft zone boundary mode justifies the analysis of the EPR data for the semiconducting samples and substantiates our assumption that the structural anomaly is related to the oxygen octahedral rotational instability.

To summarize, the phase diagram of $Sr_{1-x}Eu_xTiO_3$ has been determined experimentally by EPR and resistivity measurements with focus on the structural instability. It is found that this instability depends nonlinearly on the Eu composition x. The theoretical analysis is based on the nonlinear polarizability model and predicts a change in the dynamics around x = 0.25. The experimental phase diagram is reproduced by assuming that the double-well potential represents a doping and x dependent average of the end member potentials and by adjusting the next nearest neighbor interactions which are found to follow the x-dependence of $T_S$. From the calculations it is expected that for x ≤ 0.25 also anomalies in the acoustic mode dispersion appear which can be detected experimentally by resonant ultrasound spectroscopy and as precursor dynamics in local probe experiments, analogous to STO. The zone boundary related soft mode is found to follow mean-field behavior for all x and its temperature dependence is reflected in the EPR line width anomaly of the samples.

**Acknowledgement** This work is partly supported by the Swiss National Science Foundation, SCOPES Grant No. IZ73Z0_128242, and the Georgian National Science Foundation Grant No. GNSF/ST08/4-416.
.